\documentclass[twocolumn,secnumarabic,amssymb, nobibnotes, aps, prd,nofootinbib]{revtex4-2}

\newcommand{\m}[1]{\macro{#1}}

\usepackage{amsmath}
\usepackage{graphicx}
\usepackage{xcolor}

\renewcommand{\to}{\rightarrow}
\renewcommand{\l}{\lambda}
\def\be{\begin{equation}}
\def\ee{\end{equation}}
\def\ba{\begin{eqnarray}}
\def\ea{\end{eqnarray}}
\def\nb{\nonumber}
\def\p{\partial}

\def\a{\alpha}
\def\b{\beta}
\def\e{\epsilon}

\def\G{\Gamma}
\def\d{\delta}
\def\D{\Delta}
\def\l{\lambda}

\def\m{\mu}
\def\n{\nu}

\def\r{\rho}

\def\le{\left}

\def\pri{\prime}
\def\k{\kappa}
\def\le{\langle}
\def\re{\rangle}

\def\mc{\mathcal}

\def\ms{\mathsf}

\newcommand{\Tr}{{\rm Tr}}
\begin{document}

\title{Nucleon electric and magnetic polarizabilities in Holographic QCD}%

\author{Federico Castellani}%
\email[]{federico.castellani@unifi.it}
\affiliation{INFN, Sezione di Firenze,}
\affiliation{Dipartimento di Fisica e Astronomia, Universit\'a di Firenze,\\Via G. Sansone 1, I-50019 Sesto Fiorentino (Firenze), Italy.}
\begin{abstract}
We analyze the resonance contributions to the generalized Baldin sum rule, namely the sum of the generalized electric and magnetic nucleon polarizabilities $\a_E(Q^2)$ and $\b_M(Q^2)$, within the Holographic QCD model by Witten, Sakai, and Sugimoto (WSS). In particular, we account for the contributions from the first low-lying nucleon resonances with spin 1/2 and 3/2 and both parities.
After having extrapolated the WSS model parameters to fit experimental data on baryonic observables, we compare our findings with alternative predictions in literature, finding a qualitative agreement with all of them. Moreover, at least for the proton case, where data are available, our results are in qualitative accordance with resonance contributions extracted from experimental nucleon-resonance helicity amplitudes. 
\end{abstract}
\maketitle

\tableofcontents

\section{Introduction}
\label{sec:intro}
Nucleons compose a great part of the visible matter we are surrounded by, but despite that, our knowledge about their internal properties still needs to be completed. If on the one hand scattering processes at high energy provide a useful tool to probe the physics of the single constituents of nucleons, on the other hand, low-energy scattering processes allow us to extract crucial insights and information about the internal collective structure of neutron and proton,  described in terms of composite properties as mass, charge, spin or polarizabilities. In particular, polarizabilities quantify nucleons' response to an external electromagnetic field and the rearrangement of their internal charge, spin, or magnetization distributions. In the case in which the probing photon is virtual, \textit{i.e.} with strictly positive momentum-squared $Q^2$, polarizabilities acquire a $Q^2$ dependence and are called \textit{generalized polarizabilities}. 
The study of the low-$Q^2$ (less than 1 GeV$^2$) behavior of those observables provides a unique key to analyzing the properties and the collective dynamics of the nucleon constituents and at the same time a strict validity test of Chiral Effective Field Theory ($\chi$EFT) and Lattice QCD computations (where available). In the present work, we will focus on the electric and magnetic generalized nucleon polarizabilities $\a_E(Q^2)$ and $\b_M(Q^2)$.\\
How the generalized polarizabilities depend on $Q^2$ can be introduced in two different ways, namely through virtual Compton scattering (VCS) and forward doubly-virtual Compton scattering (VVCS) \cite{Drechsel:2002ar,Lensky:2016nui}: while in the former scattering process, the initial probing photon is virtual and the final one is real, in the latter both of them are virtual and have equal momentum. Then, it is important to stress that at finite exchanged momentum-squared a relation between the generalized polarizabilities arising from these two different types of scattering processes is not known up to now, and they are only connected in the real-photon limit \cite{Hagelstein:2015egb}.
\\From an experimental point of view, several Virtual Compton Scattering experiments have been conducted over the last twenty years, from Jefferson Lab \cite{JeffersonLabHallA:2004vsy,JeffersonLabHallA:2012zwt,Li:2022sqg}, to MAMI \cite{VCS:2000ldk,A1:2008rjb,A1:2019mrv,A1:2020nof,Blomberg:2019caf} and Bates \cite{Bourgeois:2006js,Bourgeois:2011zz}, aimed at measuring the low-$Q^2$ behavior of $\a_E(Q^2)$ and $\b_M(Q^2)$ for the proton. While some data agrees
with theoretical understanding arising from, for instance, $\chi$EFTs, no single calculation completely describes the experimental results. In particular, the most puzzling discrepancy comes from the observed low-energy behavior of the electric polarizability, which contrasts the expected predictions suggesting a monotonic decrease with increasing $Q^2$ \cite{VCS-II:2023jhu}. In particular, an unexpected peak at around $Q^2 \sim 0.3-0.4$ GeV$^2$ has been observed. It is commonly suggested, as in \cite{VCS-II:2023jhu}, that one can identify mesonic cloud effects as a possible origin of this disagreement between theory and experiments. Nevertheless, none of the current theoretical approaches, for instance, based on Heavy Baryon chiral perturbation theories (HB$\chi$PT) \cite{Bernard:1993bg}, Effective Lagrangian \cite{Vanderhaeghen:1996iz}, non-relativistic quark constituent models \cite{Liu:1996xd,Pasquini:2000ue,Pasquini:1997by} or Skyrme model \cite{Kim:1997hq}, manage to account for the non-trivial shape of  $\a_E(Q^2)$. An analysis based on first principles could involve Lattice QCD, but the high computational costs make the latter an arduous path for the moment. \\ JLab is currently planning future measurements of proton's generalized polarizabilities to catch better the shape of $\a_E(Q^2)$ with higher accuracy and reduce the currently large uncertainties in the available data on the very low-$Q^2$ behavior of $\b_M(Q^2)$ \cite{Achenbach:2023pba}.  \\
On the other hand, the analysis of the forward VVCS generalized polarizabilities is currently a hot topic as well: indeed it allows the extraction of relevant information on nucleon properties, their internal structure, and also atomic spectroscopy-related problems such as the so-called “Proton radius puzzle” (see \cite{Alarcon:2020wjg,Carlson:2015jba} for more details on the subject). Furthermore, their study is highly interested in the determination of the evolution with $Q^2$ of specific sum rules connecting dynamical nucleon properties, such as nucleon structure functions and photo-absorption cross sections, and combinations of the generalized polarizabilities themselves. Among these, in this work, we will mainly focus on the so-called generalized Baldin sum rule relating the sum of $\a_E(Q^2)$ and $\b_M(Q^2)$ to a certain integral over the Bjorken parameter of the $F_1(x,Q^2)$ structure function \cite{Drechsel:2002ar}.\\
In these directions, developments are also awaited from a theoretical point of view.  The $\chi$EFT's studies have well under control the processes involving nucleons and pion loops while partially describing processes in which nucleon resonances enter into the game. Given that the latter could play a role in the low-energy behavior of observables such as electric and magnetic polarizabilities, it is useful to develop a complementary theoretical approach to account for their effects. \\
In the present work, we study the VVCS electric and magnetic nucleon generalized polarizabilities and the Baldin sum rule at low-$Q^2$ within the Witten Sakai Sugimoto (WSS) holographic model. The latter is based on a specific $D4-D8$ setup in type IIA string theory, displaying on its gauge theory side a $SU(N_c)$ Yang-Mills theory, with $N_f$ quarks and a tower of massive adjoint fields. This theory shares with QCD relevant infrared features, such as confinement, mass gap, and chiral symmetry breaking \cite{Witten:1998zw,Sakai:2004cn}. Remarkably, the related non-perturbative physics, in the large $N_c$ limit, can be captured by a dual classical supergravity description.\\
Following closely what has been done in \cite{Bigazzi:2023odl} for the nucleon spin polarizabilities, we aim to compute in the WSS model the resonance contributions to $\a_E(Q^2)+\b_M(Q^2)$ at the low-$Q^2$ for both nucleons; we will focus on low-lying nucleon resonances, with spin 3/2 and 1/2 and both parities. To compare our findings with data, we extrapolate the WSS model parameters setting the masses of the nucleons and $\D(1232)$ resonance to their experimental values. Our results show that the $\D(1232)$ resonance gives the dominant contribution to the generalized Baldin sum rule at low-$Q^2$. Moreover, we find that $\a_E(Q^2) + \b_M(Q^2)$ in the model displays a smooth fall-off with increasing momentum squared for $Q^2 \gtrsim0.05$ GeV$^2$. However, at much smaller $Q^2$ a peak is observed (see figure \ref{abtot}): the latter may be just an artifact of the model, as it is suggested, at least for the proton case, by a comparison of our results with the expected trend for the resonances contribution obtained from helicity amplitudes data interpolation.\\ This work is organized as follows.\\
In section \ref{sec:scatproc}, we review some basic features of lepton-nucleon scattering, introducing the relevant objects for our analysis such as the hadronic tensor, the helicity amplitudes, and the generalized polarizabilities. \\In section \ref{sec:WSS}, we provide a short introduction to the Witten-Sakai-Sugimoto model of Holographic QCD, mostly focusing on the main ingredients for the holographic description of baryonic states and electromagnetic current. \\In section \ref{sec:results}, we present our results for the low-lying resonance contributions to the sum of the nucleon electric and magnetic generalized polarizabilities at low-$Q^2$. In particular, we account for positive parity $\D(1232)$, N(1440) and N(1710), and negative parity N(1535) and N(1650), resonances. Collecting all the above-mentioned contributions, we compare our findings, with chiral perturbation theory \cite{Alarcon:2020wjg,Nevado:2007dd} and MAID model predictions \cite{Drechsel:2000ct,Drechsel:1998hk}, findings from parameterization fits of the $F_1(x,Q^2)$ structure-function to experimental data \cite{Sibirtsev:2013cga,Hall:2014lea}, and with the expected behavior of the resonance contributions obtained from data interpolation of nucleon-resonance helicity amplitudes. \\
In section \ref{sec:conc}, we finally provide comments and conclusions.

\section{Nucleon internal structure}
\label{sec:scatproc}
Information on the low energy behavior of generalized nucleon polarizabilities can be experimentally extracted from the study of inclusive lepton-nucleon scattering processes
\be
\label{scattproc}
l\, N \rightarrow l'\, X\,,
\ee
where $X$ represents the unobserved hadronic final state and the lepton $l$ is usually chosen to be an electron. In particular, the dominant lepton-nucleon inelastic interaction is given by the one-virtual photon exchange with virtuality defined as
\footnote{Here we will work with a mostly plus signature metric $(-,+,+,+)$.}
\be
\label{virtuality}
q^2 =  Q^2 >0\,,
\ee
where $q^\m$ is the photon momentum. Another important kinematic variable is the Bjorken variable
\be
\label{x}
x= -\frac{Q^2}{2p\cdot q}= -\frac{Q^2}{2\n}\,, \hspace{1cm} \n = p\cdot q\,,
\ee
with $p^\m$ the target nucleon momentum. Notice that $x$ ranges between 0 and 1. \\

\subsection{Hadronic tensor and unpolarized structure functions}
The whole information on the hadronic part of the scattering process (\ref{scattproc}) is encoded in the so-called hadronic tensor (for more details see e.g. \cite{Jaffe:1996zw, Deur:2018roz, Manohar:1992tz,Drechsel:2002ar}). The latter can be expressed in terms of certain matrix elements of the electromagnetic current $J^{\mu}$ between the initial nucleon and final $X$ state as
\be
W^{\mu\nu} = \frac{1}{4\pi} \sum_{X} \langle p, s| J^{\mu}| X \rangle \langle X | J^{\nu}| p, s \rangle (2\pi)^4 \delta(p+q-p_X)\,,
\label{formula1}
\ee
where $p$ and $s$ are the target nucleon momentum and spin. 
Moreover, it is customary to rewrite the symmetric $(W^{\{\mu\nu\}})$  and anti-symmetric $(W^{[\mu\nu]})$ parts of the hadronic tensor in terms of the \textit{unpolarized} $F_1(x, Q^2)$, $F_{2}(x, Q^2)$ and  \textit{polarized} $g_1(x,Q^2)$, $g_2(x,Q^2)$ structure functions, respectively. Here we will focus on  $F_1(x, Q^2)$ and $F_{2}(x, Q^2)$ only:\footnote{See \cite{Bigazzi:2023odl} for an analysis of polarized structure functions in Holographic QCD.} actually, as it will be clear in the following, the sum of the electric and the magnetic generalized polarizabilities will be given in terms of a moment of $F_1(x, Q^2)$. \\Explicitly $(W^{\{\mu\nu\}})$ for spin $1/2$ targets reads \footnote{Electromagnetic current conservation yields that $q_\m W^{\m\n}=0$.}
\begin{multline}
\label{wsim}
W^{\{\mu\nu\}} =\left(\eta^{\mu\nu} - \frac{q^{\mu} q^{\nu}}{q^2} \right) F_1 +\\-\left[ \left(p^{\mu} - \frac{\nu}{q^2} q^{\mu}\right) \left(p^{\nu} - \frac{\nu}{q^2} q^{\nu}\right)\right] \frac{F_2}{\nu}\,.
\end{multline}
The behavior of the structure functions at low energy can not be determined using perturbative tools since notoriously QCD is strongly coupled in the infrared. Holography provides a complementary suited tool that we will employ, as in \cite{Bigazzi:2023odl}, to study the low-lying nucleon resonance contributions to $F_1(x, Q^2)$ and $F_{2}(x, Q^2)$. These contributions are encoded in the related helicity amplitudes which we introduce below. 

\subsection{Helicity amplitudes}
\label{sec:helicity}
The virtual Compton scattering processes, with baryon target $B$ and final state $B_X$ (in the present case, $B$ will be the nucleon and $B_X$ a baryonic resonance) can be described in terms of the helicity amplitudes, namely projections on the virtual photon polarization vectors of certain electromagnetic current's matrix elements (for detailed reviews see  \cite{Carlson:1998gf,Ramalho:2019ocp,Aznauryan:2008us}). More specifically, in the zero photon energy Breit frame,\footnote{The Breit frame can be defined as the frame having the photon, nucleon and final resonance spatial momenta collinear. Moreover, it is useful to further choose the frame in which energy is set to zero, $q^0 = 0$. Notice that this implies $E = E_X$.} they are given by \footnote{Here, we will choose the normalization convention in \cite{Carlson:1998gf} for the helicity amplitudes with the overall factor $1/(2m)$, where $m$ is the nucleon mass.}
\ba
\label{Gi}
G^+_{B_XB} &=&\frac{1}{2m} \big\langle B_X, h_X = + 1/2| \e^{+}_\m J^\m| B, h = -1/2 \big \rangle\,,	\nb\\
G^-_{B_XB} &= &\frac{1}{2m} \big\langle B_X, h_X = + 3/2| \e^{+}_\m J^\m| B, h =  1/2 \big \rangle\,,	\nb\\
G^0_{B_XB} &=& \frac{1}{2m} \big\langle B_X, h_X=  1/2| \e^{0}_\m J^\m| B, h =  1/2 \big \rangle\,,
\ea
where $h, h_X$ are the initial and final helicities and the virtual photon polarization vectors and momentum are given by 
\ba
\e^{\mu}_{\pm} &=& \frac{1}{\sqrt{2}}(0,\mp 1, -i,0)\,,\hspace{0.2cm} \e^{\mu}_{0} = \frac{1}{Q} (\ms q,0,0, 0)\,, \nb\\
q^\m &=& (0,0,0, \ms q)\,, \hspace{0.2cm}  \ms q = -Q\,.
\label{polvecgen}
\ea
Notice that the $ G^{-}_{B_XB}$ amplitude is non-zero only for scattering processes involving helicity $3/2$ resonances. 
Notably, the resonance contributions to $F_{1}(x,Q^2)$ and $F_{2}(x,Q^2)$ can be written in terms of helicity amplitudes as \cite{Carlson:1998gf} 
\begin{multline}
\label{carlson}
F_{1}(x,Q^2) = \sum_{m_X} \d((p+q)^{2} +m_{X}^{2})m^{2} \times\\\times\left(| G^{+}_{B_XB}|^2+| G^{-}_{B_XB}|^2\right),\\
\bigg(1+\frac{\nu^{2} }{Q^{2}m^{2}}\bigg)F_{2}(x,Q^2) = -\sum_{m_X} \d((p+q)^{2} +m_{X}^{2})\n \times \\ \times\left[ | G^{+}_{B_XB}|^2+|G^{-}_{B_XB}|^2  + 2| G^{0}_{B_XB}|^2\right]\,,
\end{multline}
where $m_X$ is the resonance mass.
In the case in which one wants to consider the more realistic possibility of having non-sharp resonances, the above expressions (\ref{carlson}) can be approximated by replacing the Dirac delta functions as follows \cite{Carlson:1998gf,Bigazzi:2023odl} 
\be
\label{notstable}
\d((p+q)^2 + m_X^2) \to \frac{1}{4\pi m_X} \frac{\G_X}{\big(\sqrt{|(p+q)^2| }- m_X\big)^2 + \G_X^2/4}\,.
\ee
Here $\G_X$ are the resonance widths, which we will simply take as inputs from experimental data \cite{ParticleDataGroup:2020ssz}.

\subsection{Electric and magnetic polarizabilities}
The collective response of a nucleon probed by an external electromagnetic field can be recast in two different sets of generalized polarizabilities: electric and magnetic ($\a_E(Q^2)$ and $\b_M(Q^2)$) and forward and longitudinal-transverse spin polarizabilities ($\gamma_0(Q^2)$ and $\delta_{LT}(Q^2)$). Here we will focus only on the formers, while for an analysis of the latter, we refer to \cite{Bigazzi:2023odl}.\\ Interestingly, from a theoretical point of view, the sum of $\a_E(Q^2)$ and $\b_M(Q^2)$ can be simply related to the structure function $F_{1}(x,Q^2)$ through a generalization to virtual photons of the so-called Baldin's sum rule \cite{Drechsel:2002ar}
\begin{multline}
\label{ab}
\a_E(Q^2) +\b_M(Q^2) = \frac{8\alpha_{em} m}{Q^4} \int_{0}^{x_0} dx\, x\,F_{1}(x,Q^2)\,.
\end{multline}
Here, $\alpha_{em}$ is the fine structure constant and $x_0$ is the so-called pion-production threshold
\be
x_0 = \frac{Q^2}{Q^2+(m_{\pi} +m)^2-  m^2}\,,
\label{ics0}
\ee
which excludes the elastic scattering contribution from the integral ($m_{\pi}$ is the pion mass).
For a fully detailed review of properties and sum rules involving generalized electric and magnetic polarizabilities, see e.g. \cite{Jaffe:1996zw, Deur:2018roz,Manohar:1992tz,Drechsel:2002ar}.\\
Using (\ref{carlson}), we can now express the sharp resonance contributions to (\ref{ab}) as follows 
\begin{multline}
\label{ab1}
\a_E(Q^2) +\b_M(Q^2) = \sum_{m_X}\big(| G^{+}_{B_XB}|^2+| G^{-}_{B_XB}|^2\big)\times \\\times \frac{8\alpha_{em} m^3}{Q^4}\int_{0}^{x_0} dx\, x\, \d((p+q)^{2} +m_{X}^{2}) \\
=  \sum_{m_X} \frac{8\alpha_{em} m^3}{( Q^2 + m_X^2-m^2)^3}\big(| G^{+}_{B_XB}|^2+| G^{-}_{B_XB}|^2\big)\,,
\end{multline}
In the more realistic case, where resonance decay widths are taken into account through (\ref{notstable}), the above result is modified as 
\begin{multline}
\label{expab}
\a_E(Q^2) +\b_M(Q^2) = \frac{2\alpha_{em}m}{\pi Q^4}\sum_{m_X} \frac{\G_X}{m_X}\times \\\times  \big[| G^{+}_{B_XB}|^2+| G^{-}_{B_XB}|^2\big] \mc F_1(Q^2, m_X, \G_X)\,,
\end{multline}
with 
\ba
\label{F1exp}
\mc F_1(Q^2, m_X, \G_X) =  \int_{0}^{x_0} dx\, \frac{x^2}{\big(\sqrt{|(p+q)^2| }- m_X\big)^2 + \G_X^2/4}\,.\nb\\
\ea

\section{Holographic QCD: the Witten Sakai Sugimoto model}
\label{sec:WSS}
The Witten-Sakai-Sugimoto (WSS) model \cite{Witten:1998zw,Sakai:2004cn} is the top-down model that provides the holographic theory closest to large $N_c$ QCD in the low-energy regime. The gauge theory describes the infrared dynamics of a specific configuration of $N_c$ $D4$-branes wrapped on a circle of radius $M_{KK}^{-1}$ and $N_f$ $D8$ branes. This is actually a $3+1$ dimensional $SU(N_c)$ gauge theory with $N_f$ quarks and a tower of adjoint matter fields, whose mass scale is given by $M_{KK}$. The theory also contains a 't Hooft-like parameter $\lambda$ measuring the ratio between the confining $SU(N_c)$ string tension and $M_{KK}^2$. In the planar, strongly coupled regime ($N_c\gg1$, $\lambda\gg1$ and $N_f\ll N_c$) the theory is holographically described by the classical supergravity solution sourced by the $D4$ branes and probed by the $D8$ ones. In particular, the effective low-energy action on the latter captures the hadronic sector of the model.
Since the $D8$ branes are wrapped on a $S^4$ sphere in the background, the related effective description (focusing on the particular $N_f= 2$ case) is given by a five-dimensional $U(2)$ Yang-Mills-Chern-Simons action of the form \cite{Sakai:2004cn}
\begin{multline}
\label{ymcs}
S_{f} = -\frac{\kappa}{2} \int{d^4xdz}\,\bigg[ \bigg(\frac12 h(z)\hat F^{\m\n} \hat  F_{\m\n} + k(z) \hat F^{\m z} \hat F_{\m z}\bigg)+\\+ 2\Tr  \bigg(\frac12 h(z) F^{\m\n} F_{\m\n} + k(z)  F^{\m z} F_{\m z}\bigg)\bigg]+
\\+ \frac{N_c}{24\pi^2} \int{\frac{3}{2}} \hat A \wedge \Tr F^2 +  \frac{1}{4} \hat A \wedge \hat F^2\,.
\end{multline}
Here, we have split the gauge field $\mc A$ in his $SU(2)$ $A^aT^a\, (a=1,2,3)$,\footnote{The $SU(2)$ generators $T^a$ are normalized as $\Tr(T^a T^b) = \frac{1}{2} \d^{ab}$.} and $U(1)$ $\hat A$ parts. Moreover, here $\mu,\nu$ are flat Minkowski indices,  $z\in(-\infty,\infty)$ is the holographic coordinate, and (with $M_{KK}=1$ units)
\be
\kappa = \frac{N_c\lambda}{216\pi^3}\,,\quad h(z) = (1+z^2)^{-1/3}\,,\quad k(z) = (1+z^2)\,,
\ee 
where the functions $h(z)$ and $k(z)$ account for the curvature of the background. \\
In the model, mesons are holographically identified with gauge field fluctuations. For instance, in the $N_f = 1$ case, these can be decomposed in terms of two complete function bases $\{\psi_n(z)\}$ and $\{\phi_n(z)\}$ as ${\mathcal A}_{\mu}(x^{\mu},z) = \sum_{n=1}^{\infty} B_{\mu}^{(n)}(x^{\mu})\psi_n(z)$ and
${\mathcal A}_{z} (x^{\mu},z) = \sum_{n=0}^{\infty} \varphi^{(n)}(x^{\mu})\phi_n(z)$, in such a way that the modes $B_\m^{(n)}$ and $\varphi^{(n)}$ get canonical mass and kinetic terms in the Minkowski part of (\ref{ymcs}). To obtain this, $\{\psi_n(z)\}$ are required to satisfy the eigenvalues equation
\be \label{eqforpsi}
-h(z)^{-1} \partial_z (k(z)\partial_z\psi_n(z))=\lambda_n \psi_n(z)\,,
\ee
while $\{\phi_n(z)\}$ are given by
\be
\phi_{n>0}(z) = \lambda_n^{-1/2}\partial_z\psi_n(z)\,, \quad \phi_0(z) = (\kappa\pi)^{-1/2}k(z)^{-1}\,.
\ee 
Then the $B_\m^{(n)}$ modes can be identified with vector and axial meson fields, for $n$ odd and even respectively. On the other hand, while for $n>0$ the $\{\phi_n(z)\}$ modes can be taken out by a redefinition of $B_\m^{(n)}$, $\phi_0(z)$ is recognized as the pion field.\\
For $N_f\geq 2$, baryons in the WSS model are represented by soliton solutions with non-trivial instanton number (coinciding with the baryon number)
\be
\label{nbclass}
n_B = \frac{1}{64\pi^2}\int{d^3xdz \, \e_{MNPQ} F^a_{MN} F^a_{PQ}}\,,
\ee
in the Euclidean subspace with coordinates $x^M\,, M = 1,2,3,z$ \cite{Sakai:2004cn,Hata:2007mb}. The construction and quantization of these types of solutions, in the large $\lambda$ and large $N_c$ regime, have been discussed in detail in the seminal works \cite{Hata:2007mb,Hashimoto:2008zw}. In the $N_f = 2$ case,  the $n_B= 1$ solution depends on a set of parameters, namely the instanton position in the 4-dimensional space $X^M$, its size $\r$,  and a set of global $SU(2)$ parameters ${\bf a}^I (I = 1,2,3,4$, with $\sum_{I} {\bf a}^{I\, 2} = 1$). Minimizing the energy of the solution classically fixes $ Z = 0$ and $\r^2 = (N_c/8\pi^2\kappa)\sqrt{6/5}$. \\
The quantum description of this solution is obtained by promoting the instanton parameters to time-dependent operators in terms of which one builds up a non-relativistic Hamiltonian, whose eigenstates, characterized by quantum numbers $\{B = (l,I_3,n_\r,n_z), s\}$, are identified with the baryonic states. 
Here, $l/2$ (with $l $ integer) fixes the spin $S$ and isospin $I$ representations (with only $I=S$ states allowed), $I_3$ and $s$ are the eigenvalues of the third components of the isospin and spin, and $n_{\rho},n_{z}$ are quantum numbers related to the operators $\rho$ and $Z$.\\ The baryon states take the form\footnote{Following conventions in \cite{Carlson:1998gf}, \\$\le B, \vec p\,^\pri, s' |B, \vec p,s\re = 2m (2\pi)^3 \d_{s^\pri s} \d^3(\vec p\,^\pri-\vec p)$.}
\ba
\vert \vec p, B, s\rangle = e^{i\vec p \cdot \vec X} R_{n_\r(l)}(\r)\psi_{n_z}(Z)|s,I_3\rangle\,,
\ea
where $R_{n_\r(l)}(\r)$ and $\psi_{n_z}(Z)$ are eigenfunctions of the $\r$ and $Z$ parts respectively of the Hamiltonian, while $|s,I_3\rangle$ contains the $SU(2)$ information of the state. For instance, the nucleon ground states are obtained by setting  $ \{B=(l= 1 ,I_3= \pm 1/2,n_\r = 0,n_z= 0), s= \pm1/2\}$, where $I_3 = 1/2$ (resp. $I_3 = -1/2$) corresponds to the proton (resp. neutron). 
Baryon properties under parity transformations are encoded by the $n_z$ quantum number: positive parity baryons have $n_z$ even, whereas negative parity ones have $n_z$ odd. For more details, we refer to \cite{Hata:2007mb,Bigazzi:2023odl,Hashimoto:2008zw}. The mass formula for the baryonic eigenstates (with $M_{KK} = 1$ and up to a subtraction of the vacuum energy \cite{Hata:2007mb}) reads
\ba
\label{massformula}
M &=& M_0 + \sqrt{\frac{(l+1)^2}{6} +\frac{2}{15}N_c^2} + \frac{2(n_\r +n_z) +2}{\sqrt{6}}\,,\nb\\M_0 &=& 8\pi^2 \k\,.
\ea
In the planar regime, $M\sim N_c$ is parametrically large, which justifies the non-relativistic approximation used in the quantization of the model.

\subsection{Electromagnetic current}
In the $N_f=2$ WSS model, the electromagnetic current, being a combination of the isoscalar $\hat J_V^{\m} $ and the isovector $J_V^{a= 3\,\m}$ components, can be built up holographically in terms of the classical baryon solution field strength as \cite{Hashimoto:2008zw}
\ba
\label{current}
J_{\mu}&=&J_{V\, \m}^{3}+ \frac{1}{N_c} \hat J_{V\, \m}\nb\\&=& -\kappa\left[k(z)\Tr(F_{\mu z}\tau^3)+\frac{k(z)}{N_c}{\widehat F}_{\mu z}\right]^{z\to\infty}_{z\to-\infty}\,.
\ea
After the above-mentioned quantization procedure, this is promoted to an operator. Its expectation values between nucleon ($B$) and resonance ($B_X$) states is given in terms of the following expressions \cite{Bigazzi:2023odl,Hashimoto:2008zw,Bayona:2011xj,Ballon-Bayona:2012txi}
\begin{widetext}
\ba
\label{BXJB}
&&\langle \hat J_V^{0}(0)\rangle_{B_X,B} =\frac{N_c}{2} \langle h_X, I_3^X|h,I_3\rangle F^1_{B_XB}(\vec q\,^2)\d_{n_{\r X}n_\r}\,,\nb\\
&&\langle \hat J_V^i(0)\rangle_{B_X,B} = \frac{N_c}{4M_0} \, \langle h_X, I_3^X| \bigg\{F^1_{B_XB}(\vec q\,^2) \big[2p^i- i \e^{ija}q_jS_a\big]+ 2q^i F^3_{B_XB}(\vec q\,^2)- F^2_{B_XB}(\vec q\,^2) (q^{i}q^{a}- \vec{q}\,^{2}\d^{ia})S_a\big]\bigg\}
|h,I_3\rangle \d_{n_{\r X}n_\r}\,,\nb\\
&&\langle J_V^{3,0}(0)\rangle_{B_X,B} = \frac{1}{4} \langle h_X, I_3^X|\langle n_{\r X}| \bigg\{F^1_{B_XtB}(\vec q\,^2) \big[4I^3 +i \epsilon^{ija}p^i q_{j}\rho^{2} \Tr  [ \tau^{3}\textbf{a}\tau^{a}\textbf{a}^{-1}]\big]+ \nb\\&&+F^2_{B_XB}(\vec q\,^2)\bigg[-M_0i q_{a}  \rho^{2} \Tr  [ \tau^{3}\partial_{0}(\textbf{a}\tau^{a}\textbf{a}^{-1})] + (\vec{p}\cdot\vec{q} q_{a} -\vec{q}^2p_{a})\rho^{2} \Tr  [ \tau^{3}\textbf{a}\tau^{a}\textbf{a}^{-1}]\bigg]\bigg\}|n_\r\rangle |h,I_3\rangle\,,\nb\\
&&\langle J_V^{3,i}(0)\rangle _{B_X,B}= \frac{M_{0}}{4}\big[i F^1_{B_XB}(\vec q\,^2) \epsilon^{ija}q_{j} + F^2_{B_XB}(\vec q\,^2)(q^{i}q^{a} - \vec{q}\,^{2}\delta^{ia})\big] \langle n_{\r X}|\rho^{2}|n_\r\rangle\,\langle h_X, I_3^X|\ \Tr [ \tau^{3}\textbf{a}\tau^{a}\textbf{a}^{-1}] |h,I_3\rangle\,,
\ea
\end{widetext}
where we used the short notation
\ba
\langle p_X,B_X, h_X| \cdot|p,B,h\rangle = \langle \cdot \rangle_{B_X\,,B}\,.
\ea
Moreover, in (\ref{BXJB}) we have defined
\ba
\label{CD}
F^1_{B_XB}(\vec q\,^2)  &=& \sum_{n =1}^{\infty}\frac{ g_{v^{n}}\le n_{zX}| \psi_{2n-1}(Z) |n_z\re}{\vec{q}\,^{2} + \lambda_{2n-1}}\,, \nb\\
F^2_{B_XB}(\vec q\,^2) &=&  \sum_{n =1}^{\infty}\frac{ g_{v^{n}}\le n_{zX}| \partial_{Z}\psi_{2n-1}(Z) |n_z\re}{\lambda_{2n-1}(\vec{q}\,^{2} + \lambda_{2n-1})}\,.
\ea
Notice that, since the WSS model exhibits the so-called vector meson dominance, all the matrix elements in (\ref{BXJB}) are written in terms of form factors (\ref{CD}) which are given in turn as functions of meson eigenfunctions $\psi_n(z)$, eigenvalues $\lambda_n$ and vector mesons decay constants
\be
\label{decconst}
g_{v^n} = -2\k \big[k(z) \p_z \psi_{2n-1}(z)\big]_{z= +\infty}\,.
\ee

\subsection{Helicity amplitudes in the WSS model}
\label{sec:helicity}
Now, having in place all the ingredients, we can attempt to holographically evaluate the helicity amplitudes (\ref{Gi}) in the Breit frame. It is important to stress that, in the model, the latter frame choice is the quite natural one in dealing with a non-relativistic quantum mechanic problem. 
In the Breit frame, using (\ref{BXJB}), the polarization vectors in (\ref{polvecgen}) and defining the operators
\be
\label{Oa}
O^a =  \Tr [ \tau^{3}\textbf{a}\tau^{a}\textbf{a}^{-1}]\,,
\ee
we get the following expressions for the $G^{\pm}_{B_XB}$ helicity amplitudes  \cite{Bigazzi:2023odl}
\begin{multline}
\label{G+}
G^+_{B_XB}=- \frac{1}{8 \sqrt 2M_0 m} \big( \d_{\eta_X,\eta}F^{1}_{B_XB}(\vec q\,^{2}) \ms q  +\\+ \d_{\eta_X,-\eta}F^{2}_{B_XB}(\vec q\,^{2}) \ms q^2 \big)\bigg[ 2\sqrt{m_Xm}\,\d_{I_X,I} \d_{n_{\r X}n_\r}+\\-i M_0^2\langle n_{\r X}|\rho^{2}|n_\r\rangle\, \langle h_X, I_3^X| \big(O_2 -iO_1\big) |h,I_3\rangle\bigg]\,,
\end{multline}
and 
\begin{multline}
\label{G-}
G^-_{B_XB} 
=\big( \d_{\eta_X,\eta}F^{1}_{B_XB}(\vec q\,^{2}) \ms q  + \d_{\eta_X,-\eta}F^{2}_{B_XB}(\vec q\,^{2}) \ms q^2 \big)\times \\ \times \frac{iM_0}{8 \sqrt 2 m} \langle n_{\r X}|\rho^{2}|n_\r\rangle\, \langle h_X, I_3^X|  \big(O_2-iO_1\big) |h,I_3\rangle\,.
\end{multline}
The explicit expressions for the matrix elements appearing in the above formulae can be found in e.g. \cite{Bigazzi:2023odl}.

\section{Numerical results: comparison with experimental data}
\label{sec:results}
In this section, we collect our results for the low-lying resonance contributions to the sum of electric and magnetic generalized polarizabilities of both nucleons at low-$Q^2$. In doing that, as it was done in \cite{Bigazzi:2023odl}, we extrapolate the WSS model parameters to realistic QCD data, namely fixing $N_c =3$ and setting the difference between the masses of the ground state nucleon and of the $\D(1232)$, as deduced from (\ref{massformula}), to its the experimental value
\ba
\label{MKK}
m_{\D(1232)} - m \approx 293\, \text{MeV}\, \quad \to \quad M_{KK}  \approx 488\, \text{MeV}\,.\nb\\
\ea
Furthermore, we fix the remaining free parameter $\lambda$ as in \cite{Fujii:2022yqh}, requiring that 
\be
\label{barparam2}
M_0 = m \approx 939\, \text{MeV}\,,\quad \to \quad \,\, \l \approx 54.4\,. 
\ee
The strategy we carry out is the following: as in \cite{Sakai:2004cn}, we get the eigenfunctions $\psi_{2n-1}(z)$ and the eigenvalues $\lambda_{2n-1}$ numerically solving the differential equation (\ref{eqforpsi}); from these, we obtain the decay constants (\ref{decconst}) and so the $F^{1,2}_{B_XB}(\vec q\,^2) $ factors in (\ref{CD}). Since we are interested in analyzing a low energy regime, we can approximate the sums in (\ref{CD}) considering only the first mesonic states in the numerical computation (actually we used the first 32 modes). Hence, from expressions (\ref{G+}) and (\ref{G-}) we can account for the helicity amplitudes describing each nucleon-resonance transition $\gamma\,B\rightarrow B_X$.\footnote{The comparison between helicity amplitudes in the WSS model and the experimental data (for the proton case) can be found in e.g.\cite{Bigazzi:2023odl,Bayona:2011xj,Ballon-Bayona:2012txi}.} Finally, using the sum rule (\ref{ab1}), we can holographically evaluate $\a_E(Q^2) + \b_M(Q^2)$ and compare this with MAID predictions and findings from alternative theoretical approaches.
\subsection{Resonance contributions to $\a_E(Q^2) + \b_M(Q^2)$ }
Here we will focus on a few low-lying resonances' contributions to $\a_E(Q^2) + \b_M(Q^2)$, following essentially similar steps to the ones of the analysis in \cite{Bigazzi:2023odl} for the evaluation of the generalized nucleon spin polarizabilities in the WSS model. In particular, we will consider the positive parity $\D(1232)$ $(l=3\,, n_{\rho}=0\,,n_z=0)$, $N(1440)$ $(l=1\,, n_{\rho}=1\,,n_z=0)$, $N(1710)$ $(l=1\,, n_{\rho}=0\,,n_z=2)$ and negative parity  $N(1535)$ $(l=1\,, n_{\rho}=0\,,n_z=1)$, $N(1650)$ $(l=1\,, n_{\rho}=1\,,n_z=1)$ resonances. Using (\ref{ab1}), (\ref{G+}), (\ref{G-}) and taking from \cite{Bigazzi:2023odl} the expressions for the expectation values of the operator (\ref{Oa}) between nucleon states and different possible resonances, we obtain that
\begin{widetext}
\ba
\label{ab2}
\left(\a_E(Q^2) +\b_M(Q^2)\right)_{X,\eta_X = 1} =&&  \frac{\alpha_{em} m_Xm^2Q^2}{4M_0^2( Q^2 + m_X^2-m^2)^3} (F^{1}_{B_XB}(\vec q\,^{2}) )^2 
\bigg[\d_{I_X,I}\bigg(\d_{n_{\r X}n_\r}+ 2I_3 \frac43 M_0^2\langle n_{\r X}|\rho^{2}|n_\r\rangle^2 \bigg)^2+\nb\\
&&+\d_{I_X,I+1}\frac{32} 9M_0^4 \langle n_{\r X}|\rho^{2}|n_\r\rangle^2 \bigg]\,,\nb\\
\ea
\ba
\label{ab22}
\left(\a_E(Q^2) +\b_M(Q^2)\right)_{X,\eta_X = -1} = && \frac{\alpha_{em} m_Xm^2Q^4}{4M_0^2( Q^2 + m_X^2-m^2)^3} (F^{2}_{B_XB}(\vec q\,^{2}) )^2 \bigg[\d_{I_X,I}\bigg(\d_{n_{\r X}n_\r} +2I_3\frac43 M_0^2\langle n_{\r X}|\rho^{2}|n_\r\rangle^2 \bigg)^2+\nb\\&&+\d_{I_X,I+1}\frac{32} 9M_0^4 \langle n_{\r X}|\rho^{2}|n_\r\rangle^2 \bigg]\,.
\ea
\end{widetext}

Here, we have distinguished between the positive and negative parity resonance contributions. Notice that the only difference between the neutron and proton cases follows from the isospin eigenvalues $I_3$ in (\ref{ab2}) and (\ref{ab22}). Now, let us analyze the neutron and proton cases separately.
 
\subsubsection{Neutron}
\begin{figure}[htb]
\begin{center}
\includegraphics*[width=0.496\textwidth]{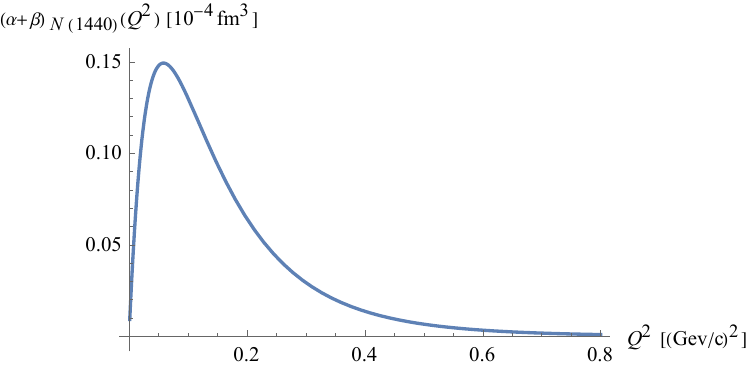}
\includegraphics*[width=0.496\textwidth]{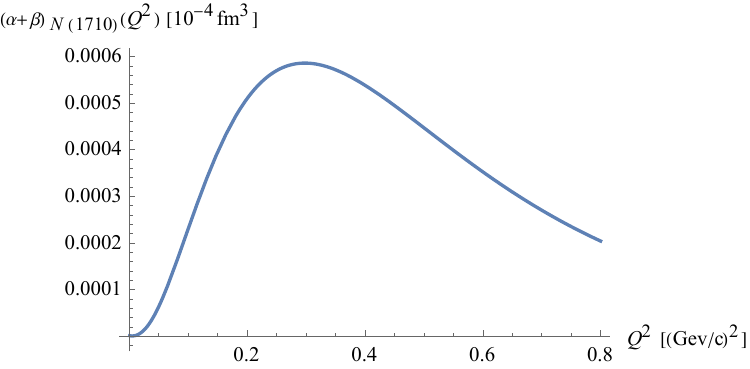}
\end{center}
\caption{\small Contributions to the neutron $\a_E(Q^2) + \b_M(Q^2)$ from sharp nucleon positive parity resonances N(1440) (top)  and N(1710) (bottom).} 
\label{abn}
\end{figure}

\begin{figure}[htb]
\begin{center}
\includegraphics*[width=0.496\textwidth]{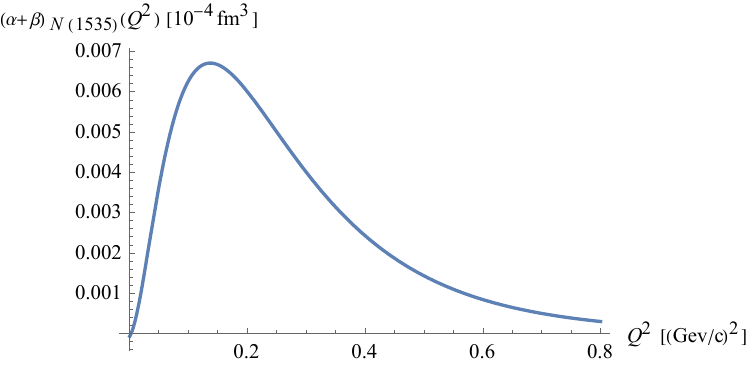}
\includegraphics*[width=0.496\textwidth]{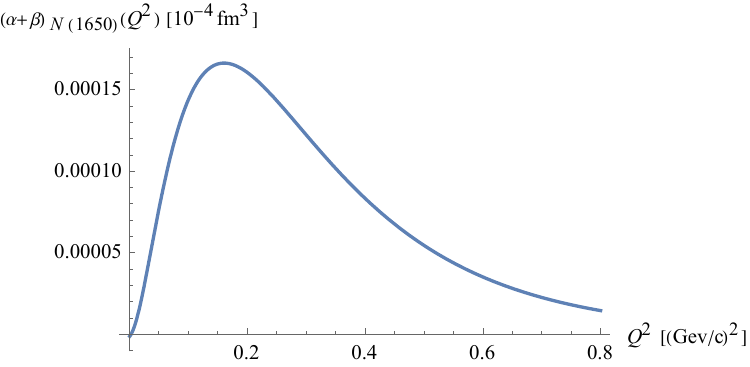}
\end{center}
\caption{\small Contributions to the neutron $\a_E(Q^2) + \b_M(Q^2)$ from sharp nucleon negative parity resonances N(1535) (top)  and N(1650) (bottom).} 
\label{abn1}
\end{figure}

\begin{figure}[htb]
\begin{center}
\includegraphics*[width=0.496\textwidth]{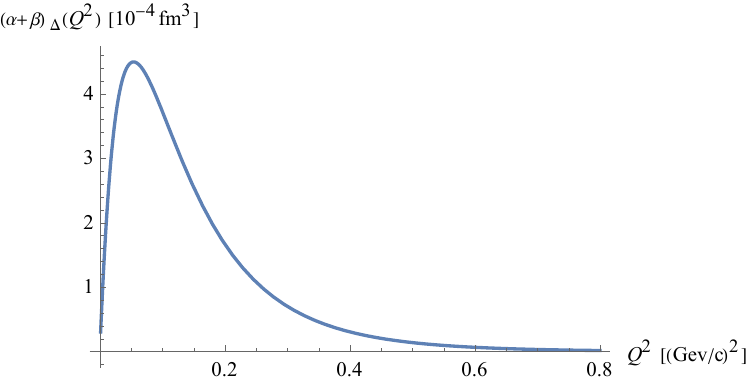}
\end{center}
\caption{\small Contribution to the neutron (and proton) $\a_E(Q^2) + \b_M(Q^2)$ from sharp nucleon $\D(1232)$ resonance.} 
\label{abnD}
\end{figure}

In figures \ref{abn} and \ref{abn1}, we show the first low-lying spin 1/2 resonance contributions to neutron's $\a_E(Q^2) + \b_M(Q^2)$ at low $Q^2$. The $\Delta(1232)$ contribution, shown in figure \ref{abnD}, is as expected the dominant one; it is isospin-independent and thus it is the same as that for the proton.  The first subdominant contribution appears to be given by the Roper resonance.

\subsubsection{Proton}
\begin{figure}[htb]
\begin{center}
\includegraphics*[width=0.496\textwidth]{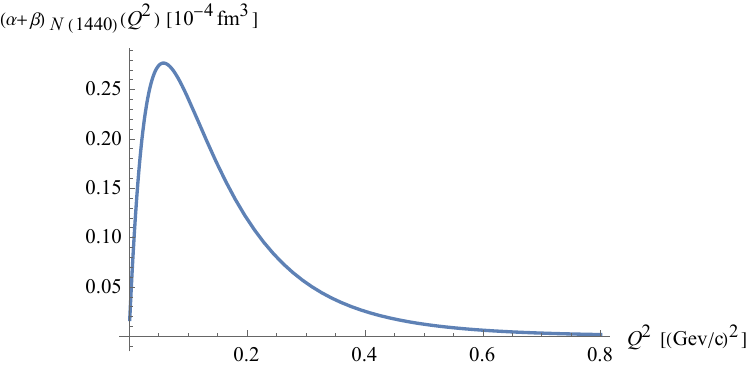}
\includegraphics*[width=0.496\textwidth]{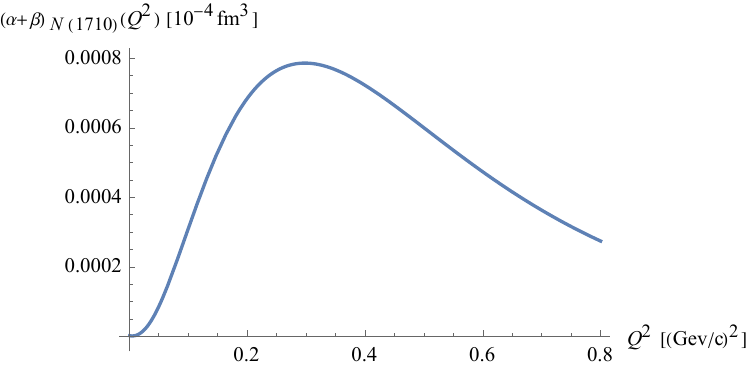}
\end{center}
\caption{\small Contributions to the proton $\a_E(Q^2) + \b_M(Q^2)$ from sharp nucleon positive parity resonances N(1440) (top)  and N(1710) (bottom).} 
\label{abp}
\end{figure}

\begin{figure}[htb]
\begin{center}
\includegraphics*[width=0.496\textwidth]{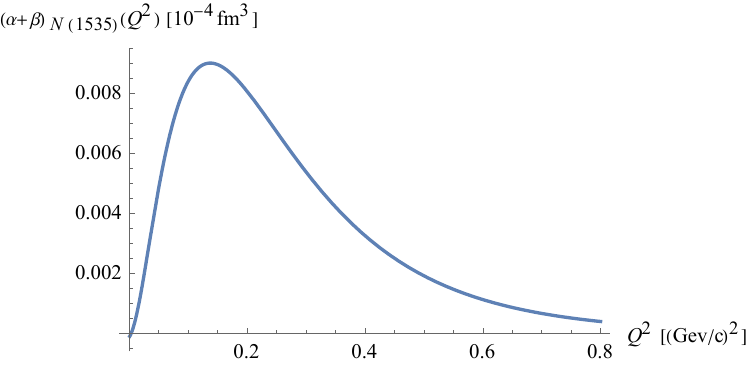}
\includegraphics*[width=0.496\textwidth]{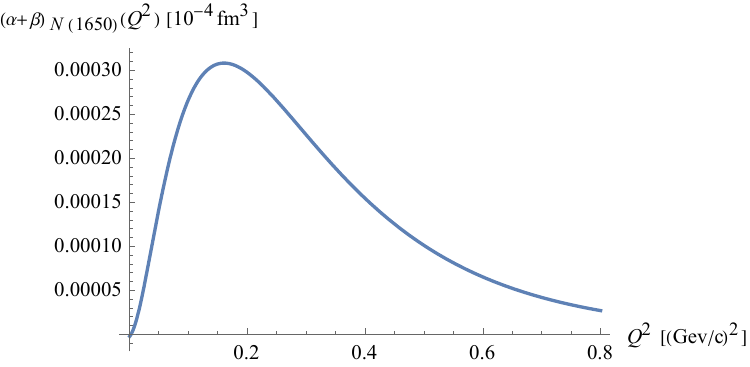}
\end{center}
\caption{\small Contributions to the proton $\a_E(Q^2) + \b_M(Q^2)$ from sharp nucleon negative parity resonances N(1535) (top)  and N(1650) (bottom).} 
\label{abp1}
\end{figure}

In figures \ref{abp} and \ref{abp1}, we show the first low-lying spin 1/2 resonance contributions to proton's $\a_E(Q^2) + \b_M(Q^2)$ at low $Q^2$. As we mentioned before, the dominant  $\Delta(1232)$ contribution is the same as the neutron one in figure \ref{abnD}. The next subdominant contribution is again given by the Roper resonance.

\subsubsection{Total resonance contribution to Baldin sum rule}
\begin{figure}[htb]
\begin{center}
\includegraphics*[width=0.496\textwidth]{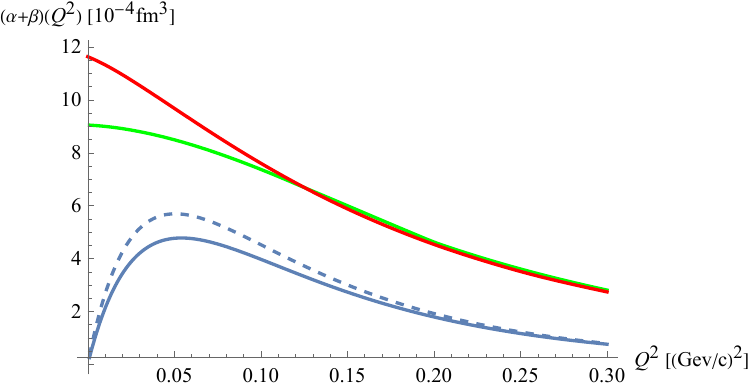}
\includegraphics*[width=0.496\textwidth]{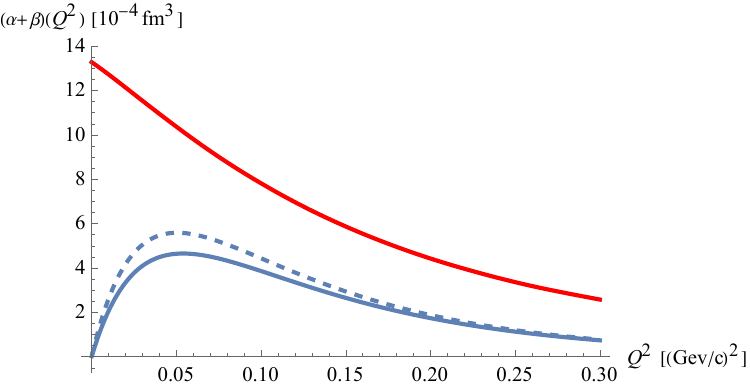}
\end{center}
\caption{\small Total contributions to the proton (top) and neutron (bottom) to the Baldin sum rule from a few low-lying nucleon resonances up to photon virtualities of $0.30$ GeV$^2$. In the proton case, 
the green solid line represents the low-energy behavior for the low-lying nucleon resonance contributions to $\a_E(Q^2) + \b_M(Q^2)$, following from the interpolation of experimental data from helicity amplitudes for $\gamma p\to B_X$ scattering processes, similarly as what has been done in \cite{HillerBlin:2022ltm,Bigazzi:2023odl}. The solid (resp. dotted) blue line represents our results for the contribution of the sharp (resp. not-sharp) resonances. Finally, the red lines corresponds to the MAID model predictions \cite{Drechsel:1998hk,Drechsel:2000ct} for the $Q^2$ evolution of the generalized Baldin sum rule.}
\label{abtot}
\end{figure}

Summing up all the contributions from the low-lying nucleon sharp resonances analyzed above we obtain the results shown with the solid blue lines in figure \ref{abtot}.\footnote{The dotted blue lines represent our predictions from the evaluation of the not-sharp resonance contributions to $\a_E(Q^2) + \b_M(Q^2)$ in (\ref{expab}).} 
Then, we can compare that with MAID model predictions \cite{Drechsel:1998hk,Drechsel:2000ct} (solid red lines), and with interpolation (solid green line) of proton helicity amplitudes data \cite{CLAS:2009ces,HillerBlin:2022ltm}. 
In particular, for the low-lying nucleon resonances, we have interpolated the experimental data on the helicity amplitudes $A_{1/2}(Q^2)$ and $A_{3/2}(Q^2)$, from which we have extracted the functions $G^+_{B_XB}(Q^2)$ and $G^-_{B_XB}(Q^2)$ respectively \cite{Bigazzi:2023odl}. Subsequently, using the latter and (\ref{ab1}), we have estimated the expected contribution of the resonances to the sum of the electric and magnetic proton spin polarizabilities.\\
As it can be seen in figure \ref{abtot}, our results, even displaying a smaller magnitude, are (at least for $Q^2\gtrsim 0.05$ GeV$^2$) in qualitative agreement with the decreasing trend shown by the above-mentioned interpolation (green line). 
Furthermore, in our approach, we find that $\a_E(Q^2)+ \b_M(Q^2)$ decreases to zero for decreasing $Q^2$, contrarily to the expected slope. Notice that this is a common feature of the resonance-nucleon helicity amplitudes and so generalized polarizabilities in the WSS model. As it was argued in \cite{Bayona:2011xj,Ballon-Bayona:2012txi} this might be due to the fact that the holographic model does not properly account for resonance decays (since in the large $\lambda$ regime baryon states are very massive and stable).
Finally, it is useful to perform a comparison with alternative theoretical predictions based for instance on chiral perturbation theory \cite{Alarcon:2020wjg} and findings deriving from empirical parametrization of the structure-function $F_1(x,Q^2)$  \cite{Sibirtsev:2013cga,Hall:2014lea}: also in these cases, we can observe a qualitative agreement on what has been found for the behavior of the generalized Baldin sum rule with $Q^2$, highlighting how in the low-energy regime the dominant contribution to polarizabilities comes from the resonance region.

\section{Conclusions}
\label{sec:conc}

In this work, we computed the nucleon resonance contributions to the sum of the electric ($\a_E(Q^2)$) and magnetic ($\b_M(Q^2)$)  nucleon generalized polarizabilities at low-$Q^2$ within the WSS holographic QCD model \cite{Witten:1998zw,Sakai:2004cn}. 
Following a similar analysis to the one in \cite{Bigazzi:2023odl}, we focused on calculating the contribution to the Baldin sum rule from the first few low-lying nucleon resonances since these are difficult to account for in other alternative theoretical approaches. More precisely, we adopted the following strategy. The electron-nucleon scattering processes by which information on the generalized polarizabilities is extracted can be described by a cross-section, given in turn in terms of the leptonic and hadronic tensors. The latter, in the case of unpolarized scattering processes, can be written as a function of the so-called nucleon structure functions $F_1(x, Q^2)$ and $F_{2}(x, Q^2)$ containing the whole information on the nucleon internal structure \cite{Jaffe:1996zw,Deur:2018roz,Manohar:1992tz}. Then, to account for the resonance contributions to $F_1(x, Q^2)$ and $F_{2}(x, Q^2)$, it is common to associate them to the experimentally accessible helicity amplitudes $G^{\pm\,,0}_{B_XB}$, related to the electromagnetic current matrix elements between the target nucleon $(B)$ and the final resonance state $(B_X)$ \cite{Carlson:1998gf,Ramalho:2019ocp,Aznauryan:2008us}. Finally, in terms of $F_1(x, Q^2)$ (or equivalently of a certain combination of $G^{\pm\,,0}_{B_XB}$) we can obtain an expression for $\a_E(Q^2) + \b_M(Q^2)$ as shown in equation (\ref{ab}).\\
 We thus evaluated the relevant helicity amplitudes and consequently the resonance contributions to the Baldin sum rule, analyzing the one-point functions of the holographic electromagnetic current between initial and final baryon states.
We have set the WSS model parameters to catch the resonance physics well, i.e. fixing the masses of the target nucleon and the lightest resonance $\D(1232)$ to their experimental values. Then, we have found that for $Q^2 \gtrsim 0.05$ GeV$^2$ the sum of the resonance contributions to $\a_E(Q^2) + \b_M(Q^2)$ displays a monotonic decreasing trend with increasing $Q^2$ (with a key role played by the kinematically preferred $\D(1232)$ contribution). \\
Both in the proton and neutron cases, our findings are in qualitative agreement with the MAID predictions  \cite{Drechsel:1998hk,Drechsel:2000ct}, chiral perturbation theory results \cite{Alarcon:2020wjg}, and finding deriving from structure-function $F_1(x, Q^2)$ parameterization fit to experimental data \cite{Sibirtsev:2013cga,Hall:2014lea}.  Moreover, similarly to the analysis in \cite{HillerBlin:2022ltm,Bigazzi:2023odl}, we compared our results for the proton with the expected behavior for the resonance contributions only to the sum of the electric and magnetic polarizabilities extracted from the experimental data interpolation of helicity amplitudes for $\gamma p\to B_X$ scattering processes, obtaining a qualitative agreement. \\
Our analysis suggests that the resonance contributions to the low-energy behavior of $\a_E(Q^2) + \b_M(Q^2)$ seem to be an essential ingredient in driving the $Q^2$ evolution of the generalized Baldin sum rule at low energy, and at the same time provides a useful solidity check for the predictions from alternative approaches based on for instance chiral perturbation theory or structure-functions empirical parametrization. Moreover, the agreement of our results with the data interpolating trend of the resonance contributions only, suggests that the WSS model, in its validity regime, manages to catch the qualitative features of the nucleon resonance physics at low energy.

\acknowledgments
We are deeply indebted to Francesco Bigazzi for suggestions, constructive discussions, and collaboration at the beginning of this project. We thank Vladimir Pascalutsa for helpful discussions and clarifications on the nucleon generalized polarizabilities.

\end{document}